\documentclass{article}
\usepackage{amsmath,amssymb,amsfonts,booktabs}
\newtheorem{thm}{Theorem}[section]
\newtheorem{prop}{Proposition}[section]

\begin{document}
\title{Takano's Theory of Quantum Painlev\'e  Equations}

\author{Yuichi UENO \\[1pt]\\(Kobe University, Japan)}
\date{}

\maketitle

{\bf Abstract}. Recently, a quantum version of Painlev\'e equations 
from the point of view of their symmetries was proposed by H. Nagoya. 
These quantum Painlev\'e equations can be written as Hamiltonian systems 
with a (noncommutative) polynomial Hamiltonian $H_{\rm J}$. We give a 
characterization of the quantum Painlev\'e equations by certain holomorphic 
properties. Namely, we introduce canonical transformations such that the 
Painlev\'e Hamiltonian system is again transformed into a polynomial 
Hamiltonian system, and we show that the Hamiltonian can be uniquely 
characterized through this holomorphic property. 

\section{Introduction}

The Painlev\'e equations $P_{\rm J}$ $({\rm J}={\rm I},\cdots ,{\rm VI})$ 
are second-order nonlinear ordinary equations without 
movable singular points. 
K. Okamoto revealed the Hamiltonian structures of the Painlev\'e equations 
and showed that there are affine Weyl group symmetries which function as 
a group of B\"acklund transformations.
In recent papers \cite{N1,N2}, H. Nagoya showed that there are quantum 
versions of Painlev\'e equations $P_{\rm II},P_{\rm III},P_{\rm IV},P_{\rm V},
P_{\rm VI}$ which have the affine Weyl group symmetries.
The relation with the KZ equation with irregular singularities is discussed by 
M. Jimbo, H. Nagoya and J. Sun\cite{JNS}, where $P_{\rm I}$ is also considered. 

In this paper, we show another construction and characterization of quantum 
Painlev\'e equations by a certain kind of the holomorphic properties. This result 
can be viewed as a quantum version of (the simpler part of) Takano's theory \cite{T,ST,MMT,M}. Our 
quantum Painlev\'e equations are given by the quantum Hamiltonian systems:
\begin{equation}\label{ttt}
\begin{array}{l}
\dfrac{df}{dt}=\dfrac{1}{h}[f,H_{\rm J}]+\dfrac{\partial f}{\partial t}\quad  $({\rm 
J=II, IV})$, \\[6pt]
\dfrac{df}{dt}=\dfrac{1}{h}[f,H_{\rm J}]+t\dfrac{\partial f}{\partial t}\quad  
$({\rm J=III, V})$, \\[6pt]
\dfrac{df}{dt}=\dfrac{1}{h}[f,H_{\rm J}]+t(t-1)\dfrac{\partial f}{\partial t}\quad   
$({\rm J=VI})$,
\end{array}
\end{equation} 
where $[ , ]$ is the commutator defined by $[q,p]:=qp-pq=h$ $(h\in \mathbb{C})$. 
The Hamiltonians $H_{\rm J}$ are as follows:
\begin{equation}\label{eq:Ham}
\begin{array}{l}
H_{\rm II}(q,p,t)=\dfrac{1}{2}p ^2-(q ^2+\dfrac{t}{2})p -bq \\[2pt]    
\phantom{H_{\rm III}(q,p,t)=}(a+b+2h=1), \\[4pt]
H_{\rm III}(q,p,t)=q ^2p ^2-q ^2p +(a+b)qp - bq + tp  \\[2pt]
\phantom{H_{\rm III}(q,p,t)=}(a+b+c+2h=1),  \\[4pt]
H_{\rm IV}(q,p,t)=tqp-qp ^2-q ^2p+ap-bq  \\[2pt]
\phantom{H_{\rm IV}(q,p,t)=}(a+b+c+h=1),  \\[4pt]
H_{\rm V}(q,p,t)=q ^2 p ^2 + tq ^2p - qp ^2 - tqp - (a+c)qp + ap + btq  \\[2pt]
\phantom{H_{\rm IV}(q,p,t)=}(a+b+c+d=1), \\[4pt]
H_{\rm VI}(q,p,t)= q^3p^2-(1+t)q^2p^2-(a+b+c)q^2p+tqp^2 \\[2pt]
\phantom{H_{\rm VI}(q,p,t)=}+(b+c+(a+b)t)qp-d(a+b+c+d-h)q-btp \\[2pt]
\phantom{H_{\rm VI}(q,p,t)=}(e=-a-b-c-2d+2h),                     
\end{array}
\end{equation}
where $a,b,c,d,e$ are parameters 
with the above relations.
We note that our resulting systems are consistent with  Nagoya's Hamiltonian 
systems.

The contents of this paper are the follwing. In section \ref{canonical},  
we give quantum versions of Takano's coordinates for the system given by equation (\ref{eq:Ham}), 
which are birational canonical transformations preserving the holomorphic of the 
system. The explicit forms of transformed Hamiltonians are given in section \ref{Ham}. 
In section \ref{chara}, we show that the system (\ref{eq:Ham}) is uniquely 
characterized by the condition on the holomorphic property in sections 
\ref{canonical},\ref{Ham}. This is the main result of this paper. In section 
\ref{affine}, we show the consistency of our result with that of H. Nagoya, which 
shows that the quantum Painlev\'e equations determined by the holomorphic have
affine Weyl group symmetry.
    
\section{Canonical transformations}\label{canonical}

In this section, we will give canonical transformations such that the 
holomorphic property of the system \eqref{eq:Ham} is preserved. They 
are explicitly given as follows.

\noindent
\textit{The case of} $P_{\rm II}$.
\begin{equation}\label{t2}
\begin{array}{ll} 
q=\dfrac{1}{x_0},\quad p=-bx_0-x_{0}^2y_0,
\quad x_0=\dfrac{1}{q},\quad y_0=-bq-q^2p, \\
q=\dfrac{1}{x_1},\quad p=t+\dfrac{2}{x_{1}^2}-ax_{1}-x_{1}^2y_{1},
\quad x_1=\dfrac{1}{q},\quad y_1=2q^4-q^2p+tq^2-aq.
\end{array}
\end{equation}

\noindent
\textit{The case of} $P_{\rm III}$.
\begin{equation}\label{t3}
\begin{array}{l}  
q=\dfrac{1}{x_0},\quad p=-bx_0-x_{0}^2y_0,
\quad x_0=\dfrac{1}{q},\quad y_0=-bq-q^2p, \\
q=x_{1},\quad p=y_{1}+\dfrac{c}{x_{1}}-\dfrac{t}{x_{1}^2}, 
\quad x_{1}=q,\quad y_{1}=p-\dfrac{c}{q}+\dfrac{t}{q^2}, \\
q=\dfrac{1}{x_2},\quad p=1-ax_2-x_{2}^2y_2,
\quad x_2=\dfrac{1}{q},\quad y_2=q^2-aq-q^2p.
\end{array}
\end{equation}

\noindent
\textit{The case of} $P_{\rm IV}$.
\begin{equation}\label{t4}
\begin{array}{l}
q=\dfrac{1}{x_0},\quad p=-\dfrac{1}{x_0}+t-cx_0-x_{0}^2y_0,
\quad x_0=\dfrac{1}{q},\quad y_0=-q^3+tq^2-q^2p-cq, \\
q=ay_{1}-x_{1}y_{1}^2,\quad p=\dfrac{1}{y_{1}},
\quad x_{1}=ap-qp^2,\quad y_{1}=\dfrac{1}{p}, \\
q=\dfrac{1}{x_2},\quad p=-bx_2-x_{2}^2y_2,
\quad x_2=\dfrac{1}{q},\quad y_2=-bq-q^2p.
\end{array}
\end{equation}  

\noindent
\textit{The case of} $P_{\rm V}$.
\begin{equation}\label{t5}
\begin{array}{l}
q=\dfrac{1}{x_0},\quad p=-t-dx_{0}-x_{0}^2y_{0},
\quad x_0=\dfrac{1}{q},\quad y_0=-tq^2-dq-q^2p, \\
q=ay_{1}-x_{1}y_{1}^2,\quad p=\dfrac{1}{y_{1}}, 
\quad x_{1}=ap-qp^2,\quad y_{1}=\dfrac{1}{p}, \\
q=\dfrac{1}{x_2},\quad p=-bx_2-x_{2}^2y_2,
\quad x_2=\dfrac{1}{q},\quad y_2=-bq-q^2p, \\
q=cy_{3}-x_{3}y_{3}^2+1,\quad p=\dfrac{1}{y_{3}},
\quad x_3=cp-qp^2+p^2 , \quad y_3=\dfrac{1}{p}.
\end{array}
\end{equation}   

\noindent
\textit{The case of} $P_{\rm VI}$.
\begin{equation}\label{t6} 
\begin{array}{l} 
q=\dfrac{1}{x_0},\quad p=-dx_{0}-x_{0}^2y_{0},
\quad x_0=\dfrac{1}{q},\quad y_0=-dq-q^2p, \\
q=1+ay_{1}-x_{1}y_{1}^2,\quad p=\dfrac{1}{y_{1}},
\quad x_{1}=ap-qp^2+p^2,\quad y_{1}=\dfrac{1}{p}, \\
q=by_{2}-x_{2}y_{2}^2,\quad p=\dfrac{1}{y_{2}},
\quad x_{2}=bp-qp^2,\quad y_{2}=\dfrac{1}{p}, \\
q=t+cy_{3}-x_{3}y_{3}^2,\quad p=\dfrac{1}{y_{3}},
\quad x_3=cp-qp^2+tp^2 ,\quad  y_3=\dfrac{1}{p}, \\
y_{0}=\dfrac{1}{y_{4}},\quad x_{0}=ex_{4}-x_{4}^2y_{4}.
\end{array}
\end{equation}

\begin{prop}
\label{prop} The system \eqref{eq:Ham} is transformed into a polynomial 
Hamiltonian system under the transformations \eqref{t2}-\eqref{t6}.
\end{prop}

The proof of this proposition is given in the next section, where we will 
give the transformed Hamiltonian in each chart explicitly.
\section{Hamiltonians on the charts}\label{Ham}

In this section, we will prove the holomorphic property (Proposition \ref{prop}). 
The proof is given by explicit computations. Since the method is similar 
in all cases, we will give the case of $P_{\rm II}$ as an example, where $x,y$ 
are used instead of $x_{i},y_{i}$:

Our $P_{\rm{II}}$ system can be written as
\begin{equation}\label{system}
\begin{cases}
\dfrac{dq}{dt}=p-q^2-\dfrac{t}{2}, \\
\dfrac{dp}{dt}=2qp+b.
\end{cases}
\end{equation}
We will transform this in terms of new coordinates given by the first 
equation in \eqref{t2}. Since $q=\dfrac{1}{x},$ we have

\begin{equation}\label{qx}
\dfrac{dq}{dt}=-\dfrac{1}{x} \dfrac{dx}{dt} \dfrac{1}{x}.
\end{equation}
From \eqref{system} and \eqref{qx}, we get

\begin{equation}
\dfrac{dx}{dt}=x^4y+(b-h)x^3+\dfrac{t}{2}x^2+1.
\end{equation}
Similarly, since $p=-bx-x^2y,$ we have

\begin{equation}
\dfrac{dp}{dt}=-b\dfrac{dx}{dt}-(x\dfrac{dx}{dt}+\dfrac{dx}{dt}x)y-x^2\dfrac{dy}{dt}.
\end{equation}
Together with \eqref{system}, we obtain

\begin{equation}
\dfrac{dy}{dt}=3(h-b)x^2y-b(b-h)x-\dfrac{t}{2}b-2x^3y^2-txy.
\end{equation}
Namely, we proved that the transformed system in the $(x,y)$ variables can be 
written again as a Hamiltonian system with the following polynomial Hamiltonian:

\begin{equation}
H=\dfrac{1}{2}x^4y^2-(h-b)x^3y+\dfrac{1}{2}b(b-h)x^2+\dfrac{t}{2}x^2y+\dfrac{t}{2}bx+y.
\end{equation}

In the same way as above, we can get Hamiltonians $H_{i}=H_{\rm 
J},_i(x_{i},y_{i},t,\alpha )$ 
on all the other charts \eqref{t2}-\eqref{t6}. The results are as follows, where $x,y$ 
are used instead of $x_{i},y_{i}$:

\noindent
\textit{The case of} $P_{\rm II}$.
\begin{equation}
\begin{array}{l}
H_{0}=\dfrac{1}{2}x ^4y^2+(b-h)x^3y +\dfrac{1}{2}b(b-h)x^2 +\dfrac{1}{2}tx^2y+\dfrac{1}{2}btx +y,  \\[8pt]
H_{1}=\dfrac{1}{2}x ^4y^2+(a-h)x^3y +\dfrac{1}{2}a(a-h)x^2 -\dfrac{1}{2}tx^2y-\dfrac{1}{2}atx -y. 
\end{array}
\end{equation}  

\noindent
\textit{The case of} $P_{\rm III}$.
\begin{equation}
\begin{array}{l}  
H_{0}=x^2y^2- tx^2y+ (-a+b-2h)xy-btx+ y,  \\ [4pt]
H_{1}=x^2y^2- x^2y + (a+b+2c)xy +(-b-c)x-ty, \\[4pt]
H_{2}=x^2y^2- tx^2y+ (a-b-2h)xy -atx-y. 
\end{array}
\end{equation}

\noindent
\textit{The case of} $P_{\rm IV}$.
\begin{equation}
\begin{array}{l}  
H_{0}=-x^3y^2 +(-a-2c+2h)x^2y+txy-c(a+c-h)x-y, \\[4pt]
H_{1}=-x^2y^3+(2a+b+2h)xy^2 -txy+x -a(a+b+h)y,  \\[4pt]
H_{2}=-x^3y^2+(-a-2b+2h)x^2y -txy-b(a+b-h)x+ y. 
\end{array}
\end{equation}

\noindent
\textit{The case of} $P_{\rm V}$.
\begin{equation}
\begin{array}{l} 
H_{0}=-x^3y^2 + x^2y^2 + (-a-2d+2h)x^2y + (a+c+2d-2h-t)xy-d(a+d-h)x +ty, \\[4pt]
H_{1}=tx^2y^3 + x^2y^2 - (2a+b+2h)txy^2 + (c-a-2h+t)xy + x +a(a+b+h)ty, \\[4pt]
H_{2}=-x^3y^2 + x^2y^2 + (-a-2b+2h)x^2y + (a+2b+c-2h+t)xy -b(a+b-h)x -ty,  \\[4pt]
H_{3}=tx^2y^3 + x^2y^2 - (b+2(c+h))txy^2 + (a-c-2h-t)xy - x +c(b+c+h)ty. 
\end{array}
\end{equation}

\noindent
\textit{The case of} $P_{\rm VI}$.
\begin{equation}
\begin{array}{l} 
H_{0}=tx^3y^2 -(1+t)x^2y^2 + (b+2d-2h)tx^2y + xy^2 \\[4pt]
\phantom{H_{0}=}+(-b-c-2d+2h-(a+b+2d-2h)t)xy \\[4pt]
\phantom{H_{0}=}+ d(b+d-h)tx +(a+b+c+2d-2h)y, \\[4pt]
H_{1}=-x^3y^4 + (2a-b-c+6h)x^2y^3 +(2-t)x^2y^2+[-a^2+a(2b+2c+d-7h)  \\[4pt]
\phantom{H_{1}=}+(b+c+d-3h)(d+2h)]xy^2+(-2a+b+c-4h+(a-b+2h)t)xy + (t-1)x  \\[4pt]
\phantom{H_{1}=}-a(b+c+d-2h)(a+d+h)y,  \\[4pt]
H_{2}=-x^3y^4 + (-a+2b-c+6h)x^2y^3 - (1+t)x^2y^2  \\[4pt]
\phantom{H_{2}=}+[-b^2+b(2c+d-7h)+(c+d-3h)(d+2h)+a(2b+d+2h)]xy^2  \\[4pt]
\phantom{H_{2}=}+(b-c+2h-(a-b-2h)t)xy-tx-b(a+c+d-2h)(b+d+h)y,  \\[4pt]
H_{3}=-x^3y^4 + (-a-b+2c+6h)x^2y^3 +(2t-1)x^2y^2  \\[4pt]
\phantom{H_{3}=}+[-c^2+cd+d^2-7ch-6h^2-dh+a(2c+d+2h)+b(2c+d+2h)]xy^2  \\[4pt]
\phantom{H_{4}=}+[-b+c+2h+(a+b-2c-4h)t]xy+t(1-t)x-c(a+b+d-2h)(c+d+h)y, \\[4pt]
H_{4}=-tx^3y^4-(3a+2b+3c+4d-10h)tx^2y^3 -(t+1)x^2y^2  \\[4pt]
\phantom{H_{4}=}-(3a^2+b^2+3c^2+5d^2+24h^2+4ab+4bc+6ca+8ad+5bd\\[4pt]
\phantom{H_{4}=}+8cd-17ah-11bh-17ch-23dh)txy^2  \\[4pt]
\phantom{H_{4}=}+(-2a-b-c-2d+4h-(a+b+2c+2d-4h)t)xy-x \\[4pt]
\phantom{H_{4}=}-(a+b+c+d-2h)(a+c+d-2h)(a+b+c+2d-2h)ty. 
\end{array}
\end{equation}

\section{Characterization of $H_{\rm J}$ by Takano's theory}\label{chara}

In this section, we characterize $H_{\rm J}$ by the holomorphic property (Takano's theory \cite{T,ST,MMT,M}).

\begin{thm}\label{thm}
In a polynomial Hamiltonian system for each variables $q,p$, 
the Hamiltonian $H_{\rm J}$ can be uniquely characterized through the holomorphic 
property under the transformations given in \eqref{t2}-\eqref{t6}.
\end{thm}

We show only the case of $J=II$, since the other cases are similar. For example, 
we first consider the case of polynomials of order 4. We parametrize such 
a general polynomial as

\begin{equation}
\begin{array}{l}
H=k_{1}q^4p^4+k_{2}q^4p^3+k_{3}q^4p^2+k_{4}q^4p+k_{5}q^4+k_{6}q^3p^4+k_{7}q^3p^3+k_{8}q^3p^2 
\\[3pt]
\phantom{H}+k_{9}q^3p+k_{10}q^3+k_{11}q^2p^4+k_{12}q^2p^3+k_{13}q^2p^2+k_{14}q^2p+k_{15}q^2+k_{16}qp^4 
\\[3pt]
\phantom{H}+k_{17}qp^3+k_{18}qp^2+k_{19}qp+k_{20}q+k_{21}p^4+k_{22}p^3+k_{23}p^2+k_{24}p
\end{array}
\end{equation}

The transformations in the first equation of \eqref{t2} are computed in a similar way as in section 
\ref{Ham}. 
Then, we find poles up to order $x^{-5}$.
Similary, for the second equation \eqref{t2}, we have poles up to order $x^{-13}$.

Solving the vanishing conditions of these residues, we have the following results 
for unknown coefficients $k_{1},\cdots ,k_{24} $:

\begin{equation}
\begin{cases}
k_{14}=-\dfrac{1}{2(a+b+2h)}, \\[2pt]
k_{20}=-\dfrac{b}{a+b+2h}, \\[2pt]
k_{23}=\dfrac{1}{2(a+b+2h)}, \\[2pt]
k_{24}=-\dfrac{1}{2(a+b+2h)}, \\[2pt]
k_{i}=0 \quad (otherwise).
\end{cases}
\end{equation}

This shows that the Hamiltonian systems with the desired holomorphic property are 
uniquely determined as follows: 
\begin{equation}
H_{\rm II}=-\dfrac{tp+2bq-p^2+2q^2p}{2(a+b+2h)}. 
\end{equation}
By normalizing the parameters as $a+b+2h=1$, we obtain equation (\ref{eq:Ham}).

The proof of Theorem \ref{thm} in the case of general degree is as follows. 
The equation for undetermined coefficients $\vec k$ is a linear inhomogeneous equation 
\begin{equation}\label{kc}
A(h)\vec k=\vec c,
\end{equation}
where the coefficients $A$ are polynomials in $h$ and the inhomogeneous term $\vec c$ 
coming from the second term in equation \eqref{ttt} is 
independent of $h$. We note that the solution of this equation 
reduces to that of the analogous problem in the classical version of Takano's theory, in the limit as $h \rightarrow 0$.
To prove the uniqueness of the solution \eqref{kc} for the generic parameter $h$, 
we need to show that ${\rm det}(A(h))$ is not identically zero. The last condition follows from the 
classical result, where ${\rm det}(A(0))\not =0$.

\section{Affine Weyl group symmetry}\label{affine}

In this section, we compare our Hamiltonian systems with the quantum Painlev\'e equations 
proposed by H. Nagoya. As a result, we find that our system is consistent 
with that of H. Nagoya, up to redefinition of parameters (and rescaling of 
canonical variables).
This means that our system has the affine Weyl group symmetry of type 
$A_{1}^{(1)},C_{2}^{(1)},A_{2}^{(1)},A_{3}^{(1)}$ and $D_{4}^{(1)}$ for 
$P_{\rm II},P_{\rm III},P_{\rm IV},P_{\rm V}$ and $P_{\rm VI}$, respectively. 

Let us recall the Hamiltonians $\hat H_{\rm J}$ $(\rm J=\rm{II},\cdots ,\rm{VI})$ 
given by H. Nagoya \cite{N1,N2} (see also \cite{JNS}).\footnote{The variables $p,q$  
correspond to $\hat p,\hat q$ in \cite{JNS}, except for the case of $P_{\rm II}$, where $p=-2\hat p,q=\hat q$.}

\noindent
\textit{The case of} $P_{\rm II}$. 
\begin{equation}\label{qp2}
\hat  H_{\rm II}=-qpq+\dfrac{1}{2}p^2-\dfrac{t}{2}p-2\alpha _{1}q,
\end{equation}
where $\alpha _{0}+\alpha _{1}=1$.

\noindent
\textit{The case of} $P_{\rm III}$.
\begin{equation}
\hat H_{\rm III}=\dfrac{1}{4}[pq(p-1)q 
+ (p-1)qpq +qpq(p-1)+q(p-1)
qp]+\dfrac{1}{2}(\alpha _{0}+\alpha _{2})
(qp+pq)-\alpha _{0}q+tp, 
\end{equation}
where $\alpha _{0}=1-2\alpha _{1}-\alpha _{2}$.

\noindent
\textit{The case of} $P_{\rm IV}$.
\begin{equation}
\hat H_{\rm IV}=-qpq-pqp+2tpq
-\dfrac{1}{2}(\alpha _{0}+\alpha _{1}-4)p-\dfrac{\alpha _{1}}{2}q
+\dfrac{1}{3}(\alpha _{0}+\alpha _{1}-4)t,
\end{equation}
where $\alpha _{0}+\alpha _{1}+\alpha _{2}=1$.

\noindent
\textit{The case of} $P_{\rm V}$.
\begin{equation}
\hat H_{\rm V}=\dfrac{1}{2}(qpqp+pqpq)
-pqp+tqpq-\dfrac{t}{2}(qp+pq)
+\alpha _{1}p+\alpha _{2}tq-
\dfrac{1}{2}(\alpha _{1}+\alpha _{3})(qp+pq),
\end{equation}
where $\alpha _{0}+\alpha _{1}+\alpha _{2}+\alpha _{3}=1$.

\noindent
\textit{The case of} $P_{\rm VI}$ (see \cite{N2})
\begin{equation}\label{qp6}
\begin{array}{l}  
t(t-1)\hat H_{\rm VI}=\dfrac{1}{6}[qp(q-1)p(q-t)
+(q-1)p(q-t)pq+(q-t)pqp(q-1) \\[3pt]
\phantom{t(t-1)\hat H_{\rm VI}}+(q-t)p(q-1)pq
+(q-1)pqp(q-t)+qp(q-t)p(q-1)] \\[3pt]
\phantom{t(t-1)\hat H_{\rm VI}}+\dfrac{1}{2}[(\alpha _{0}-1)(qp(q-1)
+(q-1)pq)+\alpha _{3}(qp(q-t)+(q-t)pq) 
\\[3pt]
\phantom{t(t-1)\hat H_{\rm VI}}+\alpha _{4}((q-1)p(q-t)
+(q-t)p(q-1))]+\alpha _{2}(\alpha _{1}+\alpha _{2})(q-t), 
\end{array}
\end{equation}
where $\alpha _{0}+\alpha _{1}+2\alpha _{2}+\alpha _{3}+\alpha _{4}=1$.

\begin{prop}\label{prop2}
The Hamiltonians \eqref{eq:Ham} and Nagoya's Hamiltonians \eqref{qp2}-\eqref{qp6} coincide, up to 
redefinitions of parameters by additive constants.
\end{prop}

We will list the relations of parameters.

\noindent 
\textit{The case of} $P_{\rm II}$.
\begin{equation}
\alpha _{1}=\dfrac{b-h}{2}.
\end{equation}
\textit{The case of} $P_{\rm III}$.
\begin{equation}
\alpha _{0}=b+h,\quad \alpha _{1}=\dfrac{1}{2}(1-a-b-2h),\quad \alpha _{2}=a+h.
\end{equation}
\textit{The case of} $P_{\rm IV}$.
\begin{equation}
\alpha _{0}=-2(-2+a+b),\quad \alpha _{1}=2(b+h).
\end{equation}
\textit{The case of} $P_{\rm V}$.
\begin{equation}
\alpha _{1}=a-h,\quad \alpha _{2}=b+h,\quad \alpha _{3}=c-h.
\end{equation}
\textit{The case of} $P_{\rm VI}$.
\begin{equation}
\begin{cases}
\alpha _{0}=1-c+h,\alpha _{1}=a+b+c+2d-h,\alpha _{2}=-d-h,\alpha _{3}=-a+h,\alpha _{4}=-b+h,  
\\[2pt]
\alpha _{0}=1-c+h,\alpha _{1}=-a-b-c-2d+h,\alpha _{2}=a+b+c+d-2h,\alpha _{3}=-a+h,\alpha _{4}=-b+h.
\end{cases}
\end{equation}

From this Proposition \ref{prop2}, we find that our system defined from the 
holomorphic property has affine Weyl group symmetry. We write down the 
symmetry transformations in the notation of Nagoya's Hamiltonian for convenience.

\noindent
\textit{The case of} $P_{\rm II}$.
\begin{center}
   \begin{tabular}{c|cc|cc} \hline
     & $\alpha _{0}$ & $\alpha _{1}$ & $q$ & $p$  \\ \hline
   $s_{0}$ & $-\alpha _{0}$ & $\alpha _{1}+2\alpha _{0}$ & 
$q+\frac{\alpha _{0}}{-p-q^2-\frac{t}{2}}$ & 
$p-q\frac{\alpha _{0}}{-p-q^2-\frac{t}{2}}-\frac{\alpha _{0}}{-p-q^2-\frac{t}{2}}q
-\frac{\alpha _{0}^2}{(-\frac{t}{2}-p-q^2)^2}$ \\
   $s_{1}$ & $\alpha _{0}+2\alpha _{1}$ & $-\alpha _{1}$ & 
$q-\frac{\alpha _{1}}{p}$ & $p$ \\ \hline
   \end{tabular}
\end{center}

\noindent
\textit{The case of} $P_{\rm III}$.
\begin{center}
   \begin{tabular}{c|ccc|ccc} \hline
     & $\alpha _{0}$ & $\alpha _{1}$ & $\alpha _{2}$ 
& $t$ & $q$ & $p$  \\ \hline
   $s_{0}$ & $-\alpha _{0}$ & $\alpha _{1}+\alpha _{0}$ & 
$\alpha _{2}$ & $t$ & $q+\alpha _{0}p^{-1}$ & $p$ \\
   $s_{1}$ & $\alpha _{0}+2\alpha _{1}$ & $-\alpha _{1}$ & 
$\alpha _{2}+2\alpha _{1}$ & $-t$ & $q$ & $p-2\alpha _{1}q^{-1}+tq^{-2}$ \\ 
   $s_{2}$ & $\alpha _{0}$ & $\alpha _{1}+\alpha _{2}$ & 
$-\alpha _{2}$ & $t$ & $q+\alpha _{2}(p-1)^{-1}$ & $p$ \\ \hline
   \end{tabular}
\end{center}

\noindent
\textit{The case of} $P_{\rm IV}$.
\begin{center}
   \begin{tabular}{c|ccc|cc} \hline
     & $\alpha _{0}$ & $\alpha _{1}$ & $\alpha _{2}$ & 
$q$ & $p$  \\ \hline
   $s_{0}$ & $-\alpha _{0}$ & $\alpha _{1}+\alpha _{0}$ & 
$\alpha _{2}+\alpha _{0}$ & $q+\frac{\alpha _{0}}{t-p-q}$ & $p-\frac{\alpha _{0}}{t-p-q}$ \\
   $s_{1}$ & $\alpha _{0}+\alpha _{1}$ & $-\alpha _{1}$ & 
$\alpha _{2}+\alpha _{1}$ & $q$ & $p+\frac{\alpha _{1}}{q}$ \\ 
   $s_{2}$ & $\alpha _{0}+\alpha _{2}$ & $\alpha _{1}+\alpha _{2}$ & 
$-\alpha _{2}$ & $q-\frac{\alpha _{2}}{p}$ & $p$ \\ \hline
   \end{tabular}
\end{center}

\noindent
\textit{The case of} $P_{\rm V}$.
\begin{center}
   \begin{tabular}{c|cccc|cc} \hline
     & $\alpha _{0}$ & $\alpha _{1}$ & $\alpha _{2}$ & $\alpha _{3}$ & 
$q$ & $p$  \\ \hline
   $s_{0}$ & $-\alpha _{0}$ & $\alpha _{1}+\alpha _{0}$ & $\alpha _{2}$ & 
$\alpha _{3}+\alpha _{0}$& $q$ & $p+\frac{\alpha _{0}}{t-q}$ \\
   $s_{1}$ & $\alpha _{0}+\alpha _{1}$ & $-\alpha _{1}$ & $\alpha _{2}+\alpha _{1}$ & 
$\alpha _{3}$ & $q+\frac{\alpha _{1}}{p}$ & $p$ \\ 
   $s_{2}$ & $\alpha _{0}$ & $\alpha _{1}+\alpha _{2}$ & $-\alpha _{2}$ & 
$\alpha _{3}+\alpha _{2}$ & $q$ & $p-\frac{\alpha _{2}}{q}$ \\ 
   $s_{3}$ & $\alpha _{0}+\alpha _{3}$ & $\alpha _{1}$ & $\alpha _{2}+\alpha _{3}$ & 
$-\alpha _{3}$ & $q-\frac{\alpha _{3}}{1-p}$ & $p$ \\ \hline
   \end{tabular}
\end{center}

\noindent
\textit{The case of} $P_{\rm VI}$.
\begin{center}
   \begin{tabular}{c|ccccc|cc} \hline
     & $\alpha _{0}$ & $\alpha _{1}$ & $\alpha _{2}$ & $\alpha _{3}$ & 
$\alpha _{4}$ & $q$ & $p$  \\ \hline
   $s_{0}$ & $-\alpha _{0}$ & $\alpha _{1}$ & $\alpha _{2}+\alpha _{0}$ & 
$\alpha _{3}$ & $\alpha _{4}$ & $q$ & $p-\frac{\alpha _{0}}{q-t}$ \\
   $s_{1}$ & $\alpha _{0}$ & $-\alpha _{1}$ & $\alpha _{2}+\alpha _{1}$ & $\alpha _{3}$ & $\alpha _{4}$ & $q$ & $p$ \\ 
   $s_{2}$ & $\alpha _{0}+\alpha _{2}$ & $\alpha _{1}+\alpha _{2}$ & 
$-\alpha _{2}$ & $\alpha _{3}+\alpha _{2}$ & $\alpha _{4}+\alpha _{2}$ & 
$q+\frac{\alpha _{2}}{p}$ & $p$ \\ 
   $s_{3}$ & $\alpha _{0}$ & $\alpha _{1}$ & $\alpha _{2}+\alpha _{3}$ & 
$-\alpha _{3}$ & $\alpha _{4}$ & $q$ & $p-\frac{\alpha _{3}}{q-1}$ \\
   $s_{4}$ & $\alpha _{0}$ & $\alpha _{1}$ & $\alpha _{2}+\alpha _{4}$ & 
$\alpha _{3}$ & $-\alpha _{4}$ & $q$ & $p-\frac{\alpha _{4}}{q}$ \\ \hline
   \end{tabular}
\end{center}

\section{Conclusions}

In this paper, we gave a construction and characterization of quantum 
Painlev\'e equations by the holomorphic properties. This may be considered as 
a first step toward the study of a "quantum Painlev\'e property". Recently, 
Y. Sasano extended Takano's theory and
discovered new equations (Sasano systems) as Hamiltonian systems with 
holomorphic \cite{S}. In particular, the series of equations having the symmetry of type of 
$D_{n}^{(1)}$ can be regarded as extensions of the Painlev\'e \rm V,\rm VI equations.  
It is natural to expect a quantum analog of Sasano's results as well.\\[4pt]

{\bf Acknowledgements.} The author would like to thank Professor Y. Yamada 
for giving helpful suggestions and encouragement, and also H. Nagoya for helpful discussions,
and Professor W. Rossman for checking English.


\begin{thebibliography}{8}
\bibitem{JNS}
Jimbo, M., Nagoya, H., Sun, J.: Remarks on confluent KZ equation for ${\frak sl}_{2}$ 
and Quantum Painlev\'{e} Equations, preprint.

\bibitem{MMT}
Matano, T., Matsumiya, A., Takano, K.: On some Hamiltonian structures of 
Painlev\'{e} systems II, J. Math. Soc. Japan, \textbf{51} (1999), 766-843.

\bibitem{M}
Matsumiya, A.: On some Hamiltonian structures of Painlev\'{e} systems III, 
Kumamoto J. Math., \textbf{10} (1997), 45-73.

\bibitem{N1}
Nagoya, H.: Quantum Painlev\'{e} systems of type $A_{l}^{(1)}$, Int. J. 
Math., \textbf{15}(10) (2004), 1007-1031.
 
\bibitem{N2}
Nagoya, H.: {\it Quantum Painlev\'{e} systems}, talk at the meeting of 
the Mathematical Society of Japan (Saitama Univ., 2007).

\bibitem{NTY}
Noumi, M., Takano, K., Yamada, Y.:  B\"acklund transformations and the 
manifolds of Painlev\'{e} systems, Funkcial. Ekvac., \textbf{45} (2002), 237-258.
 
\bibitem{S}
Sasano, Y.: Higher order Painlev\'e equations of type $D_{l}^{(1)}$, RIMS 
Kokyuroku \textbf{1473} (2006), 143-163.

\bibitem{ST}
Shioda, T., Takano, K.: On some Hamiltonian structures of Painlev\'{e} 
systems I, Funkcial. Ekvac., \textbf{40} (1997), 271-291.
 
\bibitem{T}
Takano, K.: Defining manifolds for Painlev\'{e} equations, in {\it Toward 
the exact WKB analysis of differential equations, linear and nonlinear} 
(Eds. C. J. Howls, T. Kawai and Y. Takei), 261-269, Kyoto Univ. Press, Kyoto, 2000. 
\end{thebibliography}
\end{document}